\documentstyle[preprint,aps]{revtex} 
\begin{document} 

\title{Invaded cluster algorithm for Potts models}  \author{J.
Machta, Y. S. Choi, A. Lucke and T. Schweizer}
\address{Department of Physics and Astronomy, University of
Massachusetts,  Amherst, MA 01003-3720} \author{L. M. Chayes}
\address{Department of Mathematics, University of California, Los
Angeles, CA 90095-1555 } 

\maketitle \begin{abstract} The invaded cluster algorithm, a new
method for simulating phase transitions, is described in
detail.  Theoretical, albeit nonrigorous, justification of the
method is presented and the algorithm is applied to Potts models
in two and three dimensions.  The algorithm is shown to be
useful for both first-order and continuous transitions and
evidently provides an efficient way to distinguish between these
possibilities.  The dynamic properties of the invaded cluster
algorithm are studied.  Numerical evidence suggests that the
algorithm has no critical slowing for Ising models.

\end{abstract} 

\section{Introduction} This paper discusses a new cluster method
for simulating both continuous and first-order transitions.  The
method, called the invaded cluster (IC) algorithm, was
introduced in Ref.\ \cite{MaCh95} in the context of the Ising
model.  In this paper we give a more extended discussion of the
IC method, present new data for Ising and Potts critical points
and show how to apply the method to first-order transitions in
Potts models.

The Swendsen-Wang (SW) algorithms~\cite{SwWa} and other cluster
methods~\cite{Wolff,KaDo} have led to vast improvements in the
efficiency of simulating the critical region of a variety of
spin models.  For the Potts models, these algorithms are based
on the  Fortuin-Kastelyn (FK)  \cite{FoKa} representation of the
system as a correlated bond percolation problem.   Each Monte
Carlo step in a cluster algorithm consists of generating an FK
bond configuration from the spin configuration by occupying some
of the satisfied bonds (i.e.\ bonds across which the spins are
in agreement)  of the lattice.  Clusters of connected sites are
randomly  and independently assigned a new spin value that is
the same throughout the cluster.  This creates the updated spin
configuration.   

The IC algorithm shares these basic features with other cluster
algorithms but differs in how the bond configurations are
generated.  For other cluster algorithms, satisfied bonds are
independently occupied with a  probability  that depends on the
temperature.  For the invaded cluster algorithm, bonds are
occupied in a random order until the bond configuration fulfills
a stopping condition.  For example, the stopping condition may
be a requirement on the size of the largest cluster.  For
judicious choices of stopping condition the IC algorithm
simulates the transition point of the model.

The relation between the SW algorithm and the IC algorithm is 
analogous to the relation between ordinary percolation and
invasion percolation.  In ordinary (bond) percolation
\cite{StAh}, bonds are independently occupied with probability
$p$ forming connected clusters of sites. At the percolation
threshold, $p_c$, in a large finite box, the probability that
any one of these cluster has an extent comparable to the system
size ``jumps'' from nearly zero to nearly one.   In invasion
percolation \cite{Ham,Vi,WiWi,ChKo,WiBa,ChChNe},  at least the
version most relevant to the current
work,\footnote{Conventionally, invasion percolation is
formulated as a  growth process that is initiated at a limited
number of seed sites, e.g. a single site.}  bonds are randomly
ordered and then successively occupied until a stopping
condition is reached. If, in a system of scale $L$, the stopping
rule is that some cluster is comparable in extent to $L$ then,
as is easily shown (cf.\ footnote (11) in \cite{MaCh95}) the
fraction of occupied bonds will approach $p_c$ as $L \rightarrow
\infty$.  The IC algorithm for the $q$-state Potts models can be
loosely described as the generalization of invasion percolation
to the FK random cluster models.
 
The IC algorithm has several very attractive features:  First,
the algorithm may be used to study a phase transition without
{\it a priori} knowledge of the transition temperature.  The IC
algorithm thus enjoys the property of ``uniformity'' or
``self-organized criticality.''\  \footnote{In the computer
science literature, an algorithm is called ``uniform'' if it can
be applied to problems of arbitrary size without the need for
significant precomputation. Conventional Monte Carlo sampling of
critical points is non-uniform because precomputation is
required to obtain the critical temperature.  As the system size
increases, the critical temperature must be computed to
increasing accuracy.  For the IC algorithm the critical
temperature is not used as an input.  In this setting the
concept of ``uniformity'' is akin to the idea of
``self-organized criticality.''}  The transition  temperature
is  an {\em output} of the IC algorithm just as $p_c$ is an
output of invasion percolation. In cases where the critical
temperature is unknown or not known with sufficient accuracy,
this can be a significant advantage.  Histogram
reweighting~\cite{FeSW,FeLaSw} is now the method of choice for
high precision measurements of the critical temperature.  This
method involves extrapolating from a guessed critical
temperature and its systematic errors are difficult to judge. 

The IC algorithm is also an extremely fast way to simulate
Ising-Potts critical points.  In Sec.\  \ref{secauto} we show
that autocorrelation times for the IC algorithm are
significantly smaller than for related cluster algorithms. 
Indeed, for certain quantities such as the energy and the
finite-size critical temperature, the integrated autocorrelation
time appears to approach zero as the system size increases as a
result of anti-correlations between successive Monte Carlo
steps.      

Although first devised for critical points, the IC algorithm is
also effective for studying first-order transitions. For
first-order transitions, it is convenient to have the stopping
rule control the the number of sites in the largest cluster. In
this way the average magnetization is essentially fixed and a
point in the coexistence region can be explored.  By sampling a
single point in the coexistence region the problem of
exponentially long tunneling times between phases is avoided. 
The multicanonical Monte Carlo method~\cite{BeNe} also avoids
exponential tunneling times and has some features in common with
the IC algorithm.     We also present a criterion for
distinguishing first-order from continuous transitions.  This
criterion is effective for small system sizes and easily reveals
the first-order nature of the 5-state two-dimensional Potts
model and the 3-state three-dimensional Potts model, both of
which have very weak first-order transitions.  

In finite volume, the IC algorithm does not sample the canonical
ensemble.  We  call the invariant measure sampled by the
algorithm the ``IC ensemble.''  A fundamental supposition of
this paper is that the IC ensemble is equivalent to the usual
ensembles of statistical mechanics in the thermodynamic limit.  
Just as the microcanonical and canonical ensembles agree for all
local observables in the infinite volume limit, we conjecture
that the IC ensemble agrees with these two ensembles for all
local observables.  On the other hand, global fluctuations
differ in different ensembles.  For example, energy fluctuations
are proportional to the heat capacity in the canonical ensemble
but this is not the case for the IC or microcanonical
ensembles.  Although we have no proof yet of our assertions
concerning the validity of the algorithm, in Sec.\ \ref{secjus}
we carefully state our claims and supply (non-rigorous)
arguments to back them up.

Finally, we remark that in this study we have restricted our
attention to Potts models. However, as shown by Wolff
\cite{Wolff}, it is possible to generalize the cluster methods
to a much wider class of models using an embedding procedure. 
Presumably, the same ideas should work for the IC algorithm but
these matters will not be pursued here.  Some additional spin 
models and graphical representations appropriate for the IC
algorithm are discussed  in~\cite{ChMaun}.

The rest of this paper is organized as follows, in Sec.\
\ref{secalg} the invaded cluster algorithm is described in more
detail and in Sec.\ \ref{secjus} the algorithm is justified and
compared to other simulation methods. In Sec.\ \ref{secnum}
numerical results are presented.

\section{Invaded Cluster Method for Potts models} \label{secalg}

\subsection{Potts models}

The $q$-state Potts models are defined by a collection of spins, 
$\{\sigma_i\}$ with $i$ belonging to some lattice and
$\sigma_i=1,2 \ldots q$. The Hamiltonian is given by
\begin{equation} {\cal H}= - \sum_{\langle i,j \rangle}
(\delta_{\sigma_i,\sigma_j} -1) \end{equation} where the
summation is over the bonds of the lattice.  If $q=2$, this
corresponds  to an Ising system. Here we consider hypercubic
lattices of size $N=L^d$, usually  with periodic boundary
conditions and $\langle i, j\rangle$  denotes a nearest neighbor
pair. 
	
In the seminal work of Fortuin and Kastelyn~\cite {FoKa} it was
shown that the above defined spin system gives rise to a set of
percolation-type problems known as random cluster models.  These
models are defined by weights $W$ on bond configurations
$\omega$ (collections of {\em occupied}\/ bonds) that are given
by  
\begin {equation} 
\label{eq:fkweight}
W(\omega) = p^{|\omega|}(1-p)^{|E| -
|\omega|}q^{C(\omega)}  \end{equation}  where $|\omega|$ is the
number of occupied bonds, $|E|$ is the number of bonds in the
lattice (here $dN$),  $C(\omega)$ is the number of connected
components of $\omega$  (counting isolated sites) and 
\begin{equation} \label{eq:p} p = p(\beta)= 1-e^{-\beta}
\end{equation} is the relationship between the bond density
parameter and the temperature in the spin system.  In
particular,   
\begin {equation}
 \sum_{\omega}W(\omega) = Z_\beta
\equiv \text{Tr}[e^{-\beta {\cal H}}]  \end {equation}  is the
graphical expression for the partition function in finite
volume. The weights make sense for any positive real $q$  and
the case $q = 1$ is manifestly the usual bond  percolation
problem.  In this paper we only consider positive integer values
of $q$.  It has been gradually realized, with  increasing
degrees of sophistication, 
\cite{FoKa,CoKl,AiChChNe87,ACCN,SwWa,EdSo} that the graphical
model and the spin model are ``equivalent'' e.g. expectations
of  observables in the  spin model are easily calculable in
terms of appropriate probabilities in the  random cluster
system.  From the modern perspective, the two descriptions are 
regarded as  different facets of a single larger problem that is
none  other than the annealed bond--diluted version of the
original Potts model.

In $d \geq 2$, for all $q$, there is a phase transition and for
$q = 1$ and  $2$,  this transition is continuous.  (See
\cite{Grim}, \cite {Sim} for details.)   For $q \gg 1$, it can
be proved that the transition is first-order \cite{KS}.  In two 
dimensions, it is widely accepted that the transition is
continuous for $q = 3$ and $4$ and  discontinuous for $q \geq
5$.  In $d \geq 3$, the transition is believed to be
discontinuous for all $q  \geq 3$.

\subsection{The algorithm} The invaded cluster algorithm for the
$q$--state Potts models is defined as follows:   Consider a
finite lattice on which there is some spin configuration 
$\{\sigma_i\}$. From the spin configuration, a bond
configuration is constructed via a modified form of invasion
percolation.  First, the bonds of the  lattice are assigned a
random order. Bonds, $\langle i,j\rangle$, are tested  in this
order to see if $\sigma_i = \sigma_j$.  If the latter occurs,
the bond is said to be  {\em satisfied} and it is added to the
bond configuration.  These bonds are  called  the {\em
occupied}\/ bonds.  The unsatisfied bonds are not considered for
the  remainder of the  current Monte Carlo step.  

The set of occupied bonds partitions the lattice into clusters of
connected sites.  The cluster structure evolves until a {\em
stopping condition} is  achieved. The stopping condition is
typically based on the size or topology of a single  cluster as
detailed below. When the stopping condition has been satisfied,
a new spin configuration is obtained by randomly assigning one
of the $q$ spin types to each cluster (including isolated sites)
and setting each spin in the  cluster to that value.  Statistics
are collected on the new spin/bond configuration thereby
completing a single Monte Carlo step.

Among the quantities typically measured on each Monte Carlo step
is the ratio $f$ of occupied to satisfied bonds.  As we shall
see, $f$ serves as an estimator of the temperature of the system
through the relation, $\langle f \rangle \approx 1 - e^{-\beta}$.

\subsection{Stopping rules} The IC method does not use the
temperature as an input parameter.  Instead, an appropriate
choice of stopping rules allows us to simulate the critical
point, the coexistence region of a discontinuous transition or a
point away from the phase transition.  The numerical results in
Sec.\ \ref{secnum} feature the following three stopping rules: 

\begin{itemize} \item{Extension rule}

Some cluster has maximum extent $L$ (the system size) in at
least one of the $d$ directions.  

\item{Topological rule}

Some  cluster winds around the system in at least one of the $d$
directions.  (This is the usual rule for the identification of
spanning clusters in percolation.)

\item{Mass rule}

Some cluster has at least $mN$ sites.

\end{itemize}

It is noted that for any of the above rules, the invasion stops
when exactly  one cluster satisfies the condition.  Thus, during
the evolution of the bond  configuration, when a given bond is
occupied, it is only necessary to check its cluster to see if
the  stopping condition is fulfilled.  The extension and
topological rules involve no parameters and are used to drive
the system to a phase transition point.  Both of these rules
require that some cluster is barely the size of the system. We
refer to these and related stopping rules as {\em spanning}
rules and to the unique cluster which satisfies the rule as the
{\em spanning cluster}.  

The mass rule involves an input parameter, $m$ and permits us to
simulate an arbitrary temperature and also to explore the
coexistence region of discontinuous transitions. The mass rule
is an example of a {\em fixed parameter} rule, some alternative
fixed parameter rules that we have not yet studied numerically
are: \begin{itemize}

\item{Magnetization rule}

This is closely related to the mass rule and is defined in
systems where the  spins at the boundary of the sample are all
set to 1. Thus bonds connecting the boundary sites to internal
sites may only be occupied if the latter also have spin value 1.
The stopping condition is fulfilled when the number of sites
connected to the boundary first exceeds $mN$.

\item{Susceptibility rule}

After the $k^{\rm th}$ bond has been occupied, let us suppose
that there are $r  = r(k)$ clusters, $C_1, \dots C_r$ containing
$|C_1|, \dots, |C_r|$ sites.  The stopping  condition is
fulfilled when $|C_1|^2+ \dots + |C_r|^2$ first exceeds $\chi
N$.

\item{Energy rule}

Here one counts the number of bonds whose endpoints are in the
same  connected cluster (regardless of whether the bond itself
has actually been occupied) and stops when the tally exceeds
$\varepsilon |E|$.

\item{Density rule}

Essentially equivalent to the energy rule, we simply count the 
number of bonds that have been selected and stop when the total
is $\rho|E|$.

\end{itemize}

The magnetization, susceptibility and energy rules are derived
from their thermodynamic namesakes via the FK representation. In
a system with symmetry breaking boundary conditions as described
above, the average magnetization is precisely the number of
sites connected to the boundary.   Similarly, the 
susceptibility is the average size of the bond cluster
containing the origin. In periodic boundary conditions this is
the same as the average  of the sum of the squares of the
cluster sizes divided by $N$.  Finally, (cf.\ below) the energy
per bond is related to the probability that both ends of a bond
are in the same cluster.

The mass rule and density rule both require some additional 
explanation. Let us start with the mass rule: in free or
periodic boundary conditions, the finite-volume magnetization
will always vanish by symmetry.  Presumably, this is because all
of the $q$  distinct low temperature states are equally
represented.  Below the bulk transition  temperature, this is
manifested in the FK picture by the appearance of one large
cluster and a  multitude of small clusters.  The $q$ different
values that could be assigned to the single large cluster yields
the above mentioned convex combination.  However, because of the
{\em a  priori} symmetry between these states, if we agree in
advance to always assign a fixed value to the large cluster, we
get (a finite volume approximation to) the corresponding pure
state.  Such a picture can be easily established with complete
rigor if the temperature is  low.  For $q = 2$ in $d = 2$ this
picture holds up to the critical point as can be derived from 
the main theorem in \cite{Hi,Ai} and for $q \gg 1$ it is a
consequence of the result in  \cite{BKM}.  Thus, we will suppose
that fixing the fraction of sites in the largest cluster is
equivalent to fixing the magnetization.

Turning attention to the fixed bond density rule, let us assume
that  we are in a finite box $\Lambda$  e.g.\ with periodic
boundary conditions.  The total energy  divided by the number of
bonds is given by \begin{equation} \varepsilon =
\frac1{|E|}\sum_{<i,j>}<1 
-\delta_{\sigma_i,\sigma_j}>_{\beta,\Lambda}  \end{equation}
where $<>_{\beta,\Lambda}$ denotes the thermal average at
temperature  $\beta^{-1}$ in the box $\Lambda$.  However, the
quantity $<1 -\delta_{\sigma_i,\sigma_j}>$  is easily evaluated
in the FK representation as \begin{equation} <1
-\delta_{\sigma_i,\sigma_j}>_{\beta,\Lambda} = (1-\frac 1q)(1-a)
\end{equation} where $a(q,p,\Lambda)$ is the probability that
two neighboring sites belong to the same cluster.  On the other
hand, differentiating the expression Eq. (\ref{eq:fkweight}) with
respect to $\beta$ we get \begin{equation} \varepsilon
=-\frac1{|E|}\frac{\partial\log Z}{\partial\beta}  =
-\frac1{|E|}(1-p)\frac 1Z \frac{\partial Z}{\partial p} = [1 -
\frac bp] \end{equation} where $b = b(p,q,\Lambda)$ is the
probability (in a translation and  reflection invariant system)
that a given bond is present. Evidently, there is a simple
relationship between the quantities $a$ and  $b$\footnote{If
$b_{i,j}$ is the probability that the bond $<i,j>$  is occupied
and $a_{i,j}$ is the probability that $i$ and $j$ belong to the
same cluster, the relation $(q-1)(1 - a_{i,j}) = q(1 -
b_{i,j}/p$) holds even without the assumption of translation
invariance.}. We thus argue that fixing the density of bonds has
an equivalent effect as  fixing the  density of neighboring
pairs in the same cluster. The mass and density rules have clear
computational  advantages over their counterparts, the
magnetization and energy rules respectively.  For this reason
(and certain others) these rules will supersede the  energy and
magnetization rules in the remainder of this paper.

Later we will argue that the fixed parameter rules produce, in
the  thermodynamic limit, the equilibrium system at the
temperature corresponding to the chosen  value of the
parameter.  This temperature is therefore an output of the
simulation. For first-order transitions, the goal is to find
parameter values in the ``forbidden  region''; which is in
principal easy for the magnetization and the susceptibility. 
For  continuous transitions, we may probe the critical region by
considering a sequence of magnetizations tending to zero or
susceptibilities tending to infinity.  For example, choosing
$\chi \sim N^s$ with $s < 1$ the transition temperature and the
critical behavior of local observables will again, presumably,
emerge from the output.

The fixed parameter stopping rules are related to
``constrained'' random cluster ensembles.  Consider, for
example, the {\em constrained mass ensemble} defined by
restricting attention exclusively to bond configurations
$\omega$ that satisfy the mass condition (that the largest
cluster in $\omega$ has mass $M  = mN$) and are otherwise
weighted by $q$ raised to the number of components of 
$\omega$:  \begin {equation} \label{eq:onem} W_M(\omega) =
{1\kern-0.35em 1\kern-0.30em}_{\  M}(\omega)q^{C(\omega)} \end
{equation}  where ${1\kern-0.35em 1\kern-0.30em}_{\ M}(\omega)$
is one if  $\omega$ satisfies the stated mass condition and is
zero otherwise.  Similarly, we may construct the {\em 
constrained density ensemble} and the {\em constrained
susceptibility ensemble}  using weights as in Eq.\ \ref{eq:onem}
with ${1\kern-0.35em 1\kern-0.30em}_{\  M}$ replaced by the
appropriate restricting functions.

Although we have not yet attempted a derivation, it would seem
that standard ``equivalence of ensembles'' arguments could be
developed to show that in the thermodynamic limit, these
measures are identical to the usual random cluster measures at
an appropriate value of $p$.  In this case, according to the
ideas in \cite{ACCN}, the associated measure in the spin system
converges to the corresponding Gibbs distribution.

Somewhat to our surprise, it is readily shown that the IC
density rule algorithm simulates the constrained density
ensemble.  This constitutes the current high water mark as far
as rigorous results are concerned.  We present the argument
below:\\ \\ \noindent{\bf Theorem} In finite volume, the IC
algorithm with the density rule with $\rho |E|=P$ samples the
joint bond-spin distribution defined by the weights \begin
{equation}  W(\sigma,\omega) = {1\kern-0.35em 1\kern-0.30em}_{\
P}(\omega) \Delta(\sigma,\omega).  \end {equation} In
particular, the random cluster marginal is the constrained
density ensemble defined as in Eq.\ \ref{eq:onem}.\\ \\
\noindent{\it Proof}\/ For a fixed spin configuration with at
least $P$ satisfied bonds, consider the set $\Omega_P(\sigma)$
of all bond configurations $\omega$ consistent with the spin
configuration $\sigma$ and with $|\omega|=P$:  \begin{equation}
\Omega_P(\sigma) = \{\omega\mid\  {1\kern-0.35em
1\kern-0.30em}_{\ P}(\omega) = 1 \text{ and }
\Delta(\sigma,\omega) = 1\}. \end {equation} We may define an
``algorithm'' as follows: starting from an $(\omega,\sigma)$
with ${1\kern-0.35em 1\kern-0.30em}_{\ P}(\omega)
\Delta(\sigma,\omega) = 1$, the spin moves are identical to the
usual SW or IC spin moves while the bond moves are defined by
selecting, without bias, any $\omega$ out of
$\Omega_P(\sigma)$.  It is manifestly apparent that this
``algorithm'' samples the above joint bond-spin distribution.

We claim that for the density rule, the bond moves are
equivalent to this unbiased selection from $\Omega_P(\sigma)$. 
Indeed let $\sigma$ denote a spin configuration. For the purpose
of this theorem, let us implement the IC algorithm by assigning
random numbers to the bonds and selecting the lowest $P$
satisfied bonds.  We may write $\Omega_P(\sigma)=
\{\omega_1,\ldots,\omega_R\}$ with each $\omega_k$ a subset of
the satisfied bonds of the $\sigma$ and having $|\omega_k|=P$. 
It is clear that under IC dynamics, the criterion for selecting
$\omega_k$ is that the highest bond in $\omega_k$ is lower than
the highest bond in any other $\omega \in \Omega_P(\sigma)$. 
Configurations in $\Omega_P(\sigma)$ all have the same number of
bonds, hence the probabilities are unbiased\ $\Box$.

\section{Validity of the IC Method} \label{secjus}
\subsection{Comparison to the Swendsen-Wang algorithm} The IC
algorithm is in fact very similar to the SW algorithm and this,
to  the greatest extent, is the basis of our intuition
concerning the former.  In both cases,  occupied bond clusters
are grown on top of a spin configuration by  randomly selecting
some of the satisfied bonds.  When the growth process is
stopped,  both algorithms generate the updated spin
configuration from the bond configuration by the  same procedure
(described in the second paragraph of the introduction). Thus
the ``only''  difference lies in how the bonds are selected.  

In the SW algorithm, satisfied bonds are independently accepted
with probability $p$.  If $p$ is identified with a temperature
as in the FK representation (Eq.\ \ref{eq:p}) one can show
\cite{SwWa,EdSo} that detailed balance is satisfied for both the
spins and the bonds:  For the spins, this is with respect to
the canonical Gibbs measure and  for the bonds, with respect to
the random cluster measure. The key observation \cite{EdSo} is
that the SW algorithm simulates a joint measure on spin {\em and}
bond configurations  that is defined by the weights \begin
{equation} \label{eq:joint}  W(\sigma,\omega) =
p^{|\omega|}(1-p)^{|E|- |\omega|}\Delta(\sigma,\omega). \end
{equation} In the above, $\Delta(\sigma,\omega)$ insures
consistency between the  bond and spin configurations: 
$\Delta(\sigma,\omega)$ is one if all occupied bonds in  the
configuration are satisfied, otherwise $\Delta(\sigma,\omega)$
is zero.

It is worth pointing out that the SW algorithm can in fact be
described in  the framework of an IC algorithm.  Let $S =
S(\sigma)$ denote the number of satisfied bonds in a given spin 
configuration.  Let $A_p(\sigma)$ denote the random variable
that is chosen according to  binomial statistics: \begin
{equation} {\bf P}(A_p(\sigma) = A) = (^{S}_{A})p^A (1-p)^{S-A}
\end {equation} The SW algorithm is defined by the stopping rule
that on each round, an  $A_p(\sigma)$ is drawn, and then the
growth stops as soon as the first $A_p(\sigma)$ satisfied bonds
are accepted. Viewed in this light, the SW algorithm is seen to
be quite similar to some of the IC algorithms under
consideration.  However, unlike the SW algorithm, we are unable
to write down a joint distribution similiar to Eq.\
\ref{eq:joint} for any of the IC ensembles except for the case
of the density rule.

\subsection{Away from criticality: fixed parameter algorithms}

As the title of this subsection indicates, we will provide
separate discussions of the critical and non-critical
algorithms.  In part, this is because the latter require fewer
assumptions about the equilibrium state of the Potts/FK system.
For the fixed parameter stopping rules, the fundamental
assumption is that the ensemble sampled by the IC algorithm has
the same local properties as the canonical ensemble at a
temperature that yields the chosen value for the thermodynamic
function (e.g. magnetization or susceptibility) that defines the
stopping rule. We present the argument for the susceptibility
rule though similar arguments are possible for other rules.

Consider a large Potts/FK system in equilibrium at $p <
p(\beta_c)$ (i.e. $T>T_c$). The ``system wide'' average cluster
sizes will be very close to the mean value which, in turn, is
close to the limiting infinite volume susceptibility,
$\chi(p)$.  Explicitly, for bond configuration $\omega_L$ on a
lattice of scale $L$, we may compute \begin {equation}
\label{eq:c}  C(\omega_L) = \frac{C_1^2 + \dots + C_r^2}{N} 
\end {equation} and, as $L\to\infty$, the distribution of
$C(\omega_L)$ will be sharply peaked about $\chi(p)$.

Now let us contrast the behavior of the SW and IC algorithms on
a given spin configuration. Suppose that both the SW and IC
algorithms run by assigning a random number uniformly in $[0,1]$
to each bond of the lattice.  For the IC algorithm, this
provides us with the ordering while for the SW algorithm at
parameter $p$, the instructions are to occupy all satisfied
bonds with random number less than $p$.   For the SW algorithm
we may occupy bonds as a function of continuous time $t$, $\ 0
\leq t \leq 1\ $. At time $t$, all satisfied bonds with value
less than $t$ are occupied.  Clearly, if we were to stop too
soon, e.g.\ at $t = p-\epsilon$, the value of $C$ will be, with
high probability, strictly smaller than the equilibrium value. 
Similarly if we were to go beyond $p$ to $t = p + \epsilon$, we
will get a value of $C$ that is too large.   On the other hand,
stopping, at $t = p$ as we are supposed to, yields a value of
$C$ that, by definition, is typical of the equilibrium
distribution.  

Now, starting from the {\it same} spin configuration, let us do
a step of the IC algorithm (with the susceptibility rule)
stopping when  $C = \chi= \chi(p)$. We may also envision this
operation taking place as a function of continuous time.  Now at
time $t$, the same bonds have been collected in both Monte Carlo
schemes and we can reiterate the previous discussion to conclude
that the algorithm will not stop significantly before or after
$t = p$.  If the system is in equilibrium initially then the IC
algorithm chooses nearly the same stopping time as the SW
algorithm.  Since the latter was in equilibrium, evidently under
IC, we stay in equilibrium.

The opposite perspective provides us with an equally valid
argument: Suppose it is the case (as is observed) that the
fraction $f$ of satisfied bonds that are occupied has a
distribution that is sharply peaked.  Then, each step of the IC
algorithm amounts to an iteration of the SW algorithm with
parameter value equal to $f$.  If the stopping rule demands that
$C = \chi(p)$, it follows that $f = p$.

Of course for an actual simulation in finite volume, the above
arguments are by no means a rigorous proof:  To achieve  $C =
\chi(p)$, the algorithm  stops at $ t = p + \eta_L$ where
$\eta_L$ is a random variable depending on the spin
configuration. Even if we know that $|\eta_L|\to 0$ as
$L\to\infty$, we cannot, as of yet, control the effects that
these fluctuations have on  the limiting distribution for the IC
algorithm.  Indeed, the IC distribution {\it will} differ from
the canonical distribution in finite volume.  However, it is
hard to believe that these objects  do not tend to the same
distribution in the infinite volume limit, all we lack is a
proof. Nevertheless, to summarize our argument, the assumption
that the bond/spin configurations typical of the Gibbs
distribution are close to the ones of the IC ensemble is
self--consistent in and of the fact that iterations of the IC
algorithm keep us in the vicinity of the Gibbs distribution.  

Let us now turn our attention to a discussion of the situation
when the spin configuration is not typical of the desired
equilibrium distribution; here the arguments will be somewhat
less complete.  We will again consider a given spin
configuration and compare what happens under SW vs.\ IC
dynamics. Suppose, for example, that the spin configuration is
at too high a temperature.  We may imagine that we have a
configuration that is typical of the Gibbs distribution at an
inverse temperature $\tilde\beta$ such that  $1-e^{-\tilde\beta}
\equiv\tilde p < p \equiv 1-e^{-\beta}$. If we do a single
iteration of the SW algorithm, again collecting our bonds
according to the time parameter $t$, as always, we stop when $t
= p$.  Because the temperature of the spin configuration was
higher than $1/{\beta}$, the percentage of satisfied bonds will
be relatively lower then typical for configurations that are at
the right temperature.  Evidently, when we stop, with large 
probability the average bond cluster size will be smaller than
$\chi(p)$ but larger than $\chi(\tilde p)$.  The resulting spin
configuration will therefore be of intermediate character
between those that are typical of temperatures $1/{\beta}$ and
temperatures $1/{\tilde\beta}$.   By contrast, when $t = p$, the
invaded cluster algorithm does not stop.  Hence, the fraction
$f$ of satisfied bonds that get occupied under IC dynamics is in
excess of $p$ and the new configuration (also of intermediate
character) is further towards the low temperature side.  It is
thus apparent that away from equilibrium, we are pushed in the
right direction and, as this example illustrates, we are pushed
harder in the direction of equilibrium under IC dynamics than
under SW dynamics.  Similar arguments apply to the case $p <
\tilde p$ and show that now $f < p$.  We believe, this
``negative feedback mechanism''  is ultimately responsible for
the immense reduction (or complete absence) of critical slowing
down that results from using IC dynamics.
	
Unfortunately, we now leave {\it terra firma} in order to
discuss the case where the configurations cannot be
characterized in terms of a single temperature-like parameter. 
In general, dynamically generated configurations are out of
equilibrium, however, we can consider the following proposal for
the definition of an effective temperature:  For a given
configuration $\sigma_L$ let $ C(\sigma_L,t)$ be the mean square
cluster size (as defined in Eq.\ \ref{eq:c}) observed when we
have occupied all satisfied bonds with value less than $t$.  Let
$\overline C(\sigma_L,t)$ be the average of this quantity over
all realizations of random numbers on the bonds.  A measure of
the effective temperature of the configuration $\sigma_L$ is to
compare $\overline C(\sigma_L,t)$ with $\chi(p)$, the actual
(equilibrium) susceptibility as a function of the temperature
parameter $p$.  If there is a single (non--zero) point where
these curves cross, this may be identified as an effective
temperature.  In any case, if $\overline C(\sigma_L,t=p)$ is
smaller than $\chi(p)$ we can assert that the temperature is
higher than that corresponding to the parameter $p$ while if
$\overline C(\sigma_L,t=p) > \chi(p)$ it is lower.  Obviously
the colder configurations should be heated up and the hotter
configurations should be cooled down.  For spin configuration
$\sigma$ and target susceptibility $\chi$, the typical
simulation temperature of a move will be given by $t(\chi)$ such
that $\overline C(\sigma_L,t(\chi)) = \chi(p)$.  Since the curve
$\overline C(\sigma_L,t)$ is monotone we again see the negative
feedback mechanism--if the spin configuration is colder than
that corresponding to $p$ the simulation temperature will be
higher than that corresponding to $p$ and vice versa. 
 
Of course these considerations apply as well to the other fixed
parameter IC algorithms where we would argue--just as
persuasively--that the magnetization vs. $t$ or energy vs. $t$
profiles of a spin configuration can be used to measure an
effective temperature. It is worth remarking that there is one
system  where these suppositions are exact, namely the
long-ranged mean-field Ising model.  Here, for an $N$ site
system, each site interacts with every other site via a coupling
that scales inversely with $N$. At non-critical temperatures,
the SW algorithm drives this system to equilibrium exponentially
quickly (as expected) while, by contrast, the IC algorithm
achieves equilibrium in at most two Monte Carlo
steps~\cite{ChMaun}.

\subsection{At criticality: spanning algorithms} \label{valspan}
The general philosophy that underpins our belief in the validity
of the critical algorithms is quite similar to the non--critical
cases.  The important differences lie in the implicit
assumptions we have made concerning the behavior of the
graphical representation at criticality --in particular in
finite volume.  Indeed, for a Potts system with a continuous
transition,  if we ask for the value of the parameter where the 
susceptibility is equal to a certain value, the answer is
unambiguous.  Furthermore, provided that $L$ is large compared
to the typical size of bond clusters, the statistics in a finite
lattice of side $L$ should represent an excellent approximation
to the infinite volume behavior.

For the critical algorithms, the entire premise begins with
non-trivial questions about the equilibrium critical behavior 
of the random cluster model in finite volume. For example, in a
large system is there a single cluster with the following two
properties? \begin{enumerate} \item The extent of the cluster is
the scale of the system. \item The cluster does not contain two
(or more) disjoint subsets each of which satisfy condition (1).
\end{enumerate}

\noindent  Under the standard assumptions concerning the nature
of the critical point  \footnote{These include absence of an
intermediate phase, uniqueness of the phase above and at $T_c$,
uniqueness of the $q$ magnetized phases among the
translationally invariant states below $T_c$, continuously
diverging correlation lengths using several definitions of the
correlation length and positivity of the surface tension below
$T_c$.  Nearly all of these properties have been established for
independent percolation and a significant (but not sufficient)
fraction in the Ising case.}   the following picture, in the
graphical representation, emerges:  Above $T_c$, the probability
that there is any such cluster goes to zero exponentially in the
scale of the system.  Below $T_c$, there is a single large
cluster that exhausts a fraction--equal to  the spontaneous
magnetization--of the system.   Within this cluster, there are
many separate paths that are the scale of the system.  To
prevent  this large cluster from happening, requires a
fluctuation presumably as rare as $\exp [-\text{const.}L^{d-1}]$.
By process of elimination, the only place that a cluster with
the above properties could exist on all scales is the critical
point.  

This is not to say that the above line of reasoning proves that
these or other kinds of spanning clusters are indeed  typically
observed at the critical point (e.g.\ they could be power law
rare) however, as we will see, the full validity of such an
assertion is not essential for the broader features of our
argument. 

Indeed, the basic reasoning is now pretty much the same as in
the non-critical case:  If the distribution of $f$ (the fraction
of satisfied bonds that are occupied) is sharply peaked, the
central value must correspond to that of the critical parameter,
$p(\beta_c)$.  If this central value were too small then the 
correlation length would be a small fraction of the system and
the clusters would not get big enough. If the peak value is so
large that the (low temperature) correlation length is small
compared to the system size, the biggest cluster would be too
big.  Thus, the value of $f$ has to be close enough to the
critical point to ensure that the correlation length is at least
a scaling fraction of the system size.

If the required spanning cluster is itself, somehow, atypical of
criticality, the above  argument is still valid.  As a concrete
example, consider a stopping rule that  terminates the cluster
growth when there is a cluster of size  $\sqrt L$.  Such a bad
choice of a spanning cluster will nonetheless heavily favor the
critical value of $f$  over any non--critical value.  The worst
that could happen is that the scaling   of the spanning cluster
itself might be of the wrong type but in any case all  local
observables will still take on their critical values.  Finally,
starting  at non--critical spin configurations, the negative
feedback mechanism discussed  in the previous subsection applies
to these algorithms as well.

The weaker point in our argument concerns our reasoning as to why
the distribution of the $f$ values should be sharply peaked. 
(First and foremost, this has been observed in every critical
system.) Let us imagine the problem in an infinite volume
setting and consider a critical spin configuration.  We again
regard the process of growing the clusters as an independent
percolation problem defined on the random graph that is provided
by the satisfied bonds of the spin configuration.  Let us first
assume that, in the usual sense, this problem has a sharp
percolation threshold, $t_c$.  It then follows easily that $t_c$
corresponds to the critical value of the FK parameter;  $t_c = 1
- e^{\beta_c}\equiv p(\beta_c)$.  Indeed, at $t = p(\beta_c)$,
we have achieved the critical FK bond configuration and our
assumption of a sharp threshold rules out the possibility of any
other value. Going back to finite volume and starting from a
critical configuration, the argument in this case is finished: 
If we stop at an $f$ significantly different from $p(\beta_c)$,
we will get the wrong sort of clusters and  stopping at
$f\approx p(\beta_c)$ we keep the spin configuration critical.

Unfortunately, in $d=2$, it is not the case that the underlying
percolation problem has a sharp transition. Specifically, in the
Ising model on the two-dimensional square lattice it was shown
in~\cite{CNPR,CoNaPeRu} that percolation of one spin type is
necessary and sufficient for the  existence of a positively
magnetized phase.  (The analog of this  result for the general
$q$--state Potts models was proved in~\cite{Chun})

In particular, this means that in a critical configuration, there
is no infinite cluster of satisfied bonds and thus, even if
$t=1$, there is no percolation in our secondary process. 
Evidently the percolation clusters on the critical spin
configuration will themselves look critical for all $t$ between
$p(T_c)$ and 1.  It is easy to believe (but hard to prove so we
will spare the reader the details of the heuristics) that the
clusters will not go critical until $t = p(T_c)$.  Thus, the
algorithm will not stop collecting bonds until at least this
point.  However, in finite volume, there are, undoubtedly,
typical critical spin configurations  that forbid the existence
of a spanning cluster. \footnote{For example, if there are
star-connected chains (meaning that neighbors and next nearest
neighbors both count as connected) of plus spins and of minus
spins winding both ways around the torus, the topological
condition cannot be satisfied.  At the critical point, such
configurations presumably have uniformly positive probability on
all scales.} Of course this kind of disaster is ruled out by the
mechanics of the algorithm:  whatever the stopping condition, if
it was satisfied on the last iteration, it is satisfiable on the
next one.  However, near disasters can occur causing
``bottlenecks''--situations where one of a relatively few bonds
{\it must} be occupied in order to achieve a spanning cluster. 
This would have a tendency to drive us to higher values of $f$.  

We believe that these bottlenecks do occur and, in fact, are
responsible for the relatively broad tails in the distribution
of  $f$ in the region $p(\beta_c) < f < 1$ that have been
observed in our two-dimensional simulations.  However, we also
believe that these events affect only the details of how the
$L\to\infty$ limit is achieved, not the limit itself since there
are alternate routes circumventing bottlenecks occurring on all
scales.   Nevertheless, the finite-size scaling is sometimes
quite complicated and, in certain instances, we must resort to
semi-empirical fitting of the data.

An interesting feature of the IC algorithm (using spanning
rules) is that the approach to equilibrium is along a critical
trajectory.  For example, if the starting configuration is
characteristic of zero temperature, the initial bond
configuration is typical of ordinary bond percolation at
threshold. Thus some sort of power law correlations are actually
established on the first step.

\section{Numerical methods and results} \label{secnum}
\subsection{Implementation of the algorithm} The most difficult
part of the IC algorithm  is the construction of a cluster
configuration from a spin configuration.  The first step is to
produce a random permutation of the set, $E$ of bonds of the
lattice ($|E|=dL^d$ here).  This is accomplished through $|E|$
random pairwise permutations.  Initially, let
$\pi:\{1,\ldots,|E|\} \rightarrow E$  be some conventional
initial order on $E$.  For $j=1$ to $|E|$, $\pi$ is updated by
choosing a random number, $r$ in the range $j$ to $|E|$; then
the $j^{\rm th}$ and $r^{\rm th}$ elements of the permutation
are interchanged, $\pi(j) \rightleftharpoons \pi(r)$.  It is
well known that after $|E|$ steps, $\pi$ is a random
permutation.  The computational work involved in making the
random permutation is nearly linear in the number of bonds.

Bonds are explored in the order given by the random
permutation.  If a bond is satisfied it is added to the cluster
configuration.  The data structure for the cluster configuration
and its updating is done in the same general way as for other
cluster and percolation algorithms using the Hoshen-Kopelman (or
``disjoint set forests'') method.  Each cluster of sites is
described as a rooted tree and when two clusters are combined,
the root of the smaller cluster becomes a son of the root of the
larger cluster.  When two clusters of the same size are combined
the conventional direction associated with the bond that joined
the clusters determines which site is to be the root. 
Information concerning the current state of the cluster as a
whole, such as its mass, is stored with the root. 

After the cluster configuration is updated by the addition of a
bond it is necessary to check whether the current cluster
satisfies the stopping rule.  For the fixed parameter rules this
is straightforward.  For the other spanning rules, it is
necessary to associate with each site a vector from the site to
the root of its cluster.  The set of distance vectors, $\{v_i\}$
is updated in the natural way when two cluster are combined. 
The sites in the larger cluster retain their previous
coordinates relative to the root. The sites in the smaller
cluster take new coordinates, $v^{\prime}$,  \begin{equation}
v^{\prime}_i \leftarrow v_i - v_k - e_{jk} + v_j \end{equation}
where $e_{jk}$ is the unit vector of the new bond added to the
lattice, $j$ is the site in the larger cluster which connects to
$k$ in the smaller cluster.  For the topological rule,  stopping
can only occur if the new bond is added as an internal bond.  If
the new bond $\langle j,k \rangle$ is an internal bond we
evaluate,  \begin{equation} v^*_k \leftarrow v_j-e_{jk}. 
\end{equation}   If $v^*_k\neq v_k$ the current cluster is
multiply connected and the topological rule is satisfied.  

For the extension rule, each cluster must have associated with
it the coordinates of the $2d$ sites which are the most distant
from root along the $d$ axes in the positive and negative
directions. Updating these coordinates after two clusters are
combined is somewhat tedious due to periodic boundary conditions
and is described in \cite{Luckethesis}.

The invaded cluster algorithm requires 1.8$\times 10^{-5}$
seconds per update per spin on a DEC Alpha 2100 workstation. 
The running speed is about a factor of two slower than for the
SW algorithm.

For the results reported here we start with an initial ordered
($T=0$) configuration.  Unless otherwise stated, the system is
allowed to equilibrate for two hundred Monte Carlo steps (MCs)
before data collection.  If no error bars are shown in a figure,
the error is smaller than the symbol size.

\subsection{Continuous transitions}

\subsubsection{Three-dimensional {I}sing model}
 
Figure \ref{fig:spancompare} shows data for the mean value of the
ratio of occupied to satisfied bonds, $\langle f \rangle$ for the
three-dimensional Ising model as a function of $L^{-1.59}$. The
power of $L$ is chosen to approximate the inverse of the
three-dimensional Ising correlation length exponent $1/\nu
\approx 1.59$.  Results for both the topological and extension
spanning rules are shown. Finite-size corrections are smaller
for the topological stopping rule. The best linear fit to the
data for the topological rule yields 0.35803 in comparison with
a recent value $p(T_c)=$0.358098(7) from Ref.\ \cite{FeLa}. 

Using ideas made plausible by finite-size scaling theory we can
obtain two independent critical exponents.  Figure
\ref{fig:fdiff} shows  $\log(\langle f \rangle_t - \langle f
\rangle_e)$ plotted against $\log(L)$ for the three-dimensional
Ising model where $\langle f \rangle_t$ is measured using the
topological rule and $\langle f \rangle_e$ is measured using the
extension rule.  A fit to the data yields $\nu=.63$, which
is in agreement with the value 0.6289(8) from \cite{FeLa}. A
second independent exponent can be obtained either from the
cluster size distribution of the scaling of the largest
cluster.  Figure \ref{fig:csdist3d-2} shows $n(s)$, the number
of clusters per site of size $s$ with the data binned in
octaves.  This figure shows that the cluster size distribution
is indeed self-similar and allows us to estimate the exponent
$\tau$, defined by $n(s) \sim s^{-\tau}$.  A straight line fit
yields $\tau=2.19$ compared to the accepted value of 2.21.  A
more efficient way to obtain a second independent exponent is
via the fractal dimension of the spanning cluster.  The average
size of the spanning cluster is plotted against the system size
in the inset of Fig.\ \ref{fig:csdist3d-2} from which we obtain
$\beta/\nu =2.45$ compared to the accepted value 2.47.

\subsubsection{Critical two-dimensional Potts models}

Figure \ref{fig:f2dising} shows results for the extension rule
applied to the two-dimensional Ising model.  Both the mean and
median value of $f$ are plotted against $L^{-1}$ (in accord with
finite-size scaling since $\nu=1$ for the two-dimensional Ising
model) and are seen to converge to the exactly known value of
$p(T_c)$.  The fact that median lies below the mean shows that
the distribution is skewed toward larger values of $f$.  This is
presumably caused by the simultaneous percolation of the spins
discussed in Sec.\ \ref{valspan}.  This is in contrast to the
three-dimensional Ising model for which the $f$ distribution is
very symmetrical.  The inset to Figure \ref{fig:f2dising} shows
var$(f)^{1/2}$ plotted against $1/L$.  The solid line is a fit
to the data whose leading behavior is $L^{-1/2}$.  This curve
supports the hypothesis that the distribution of $f$ becomes
sharp as $L \rightarrow \infty$.

Figure \ref{fig:energypm2d-2} is a plot of the average energy per
spin\footnote{In this section, $\varepsilon$ refers to the energy
per spin rather than the energy per bond.} vs.\ $L^{-1}$ with the
exact value plotted on the vertical axis.  A fit to the data of
the form,
$\varepsilon_{0}+\varepsilon_{1}L^{-1}+\varepsilon_{2}L^{-2}$
yields, $\varepsilon_{0}=-1.706$ which is reasonably close to
the exact value $-1.7071\ldots$.  Energy fluctuations are shown
in the inset of Fig.\ \ref{fig:energypm2d-2}. The quantity
var$(\varepsilon)N$ is seen to increase roughly linearly in
$L$.  This is in contrast to the canonical ensemble where
var$(\varepsilon)N$ is the specific heat and diverges
logarithmically in $L$ for the two-dimensional Ising model.  The
behavior of energy fluctuations underscores the difference
between the IC ensemble and the canonical ensemble.

The top panel of Fig.\ \ref{fig:pm2d-2} shows the fraction of
occupied bonds vs.\ the mass of the largest cluster for the mass
rule.  Data collapse for a range of system sizes predicted by the
finite-size scaling ansatz,  \begin{equation}  \label{eq:fssW}
[\langle f \rangle-p(T_c)]L^{1/\nu} \sim G(mL^{\beta/\nu}).
\end{equation}   is confirmed in the lower panel. These results
demonstrate that the IC algorithm can be used to extract
quantitative results for the critical temperatures and critical
exponents using systems of modest size.  

Figure \ref{fig:qpottf} is a log-log plot of the deviation of
$f$ from its exact value versus the system size for
two-dimensional Potts models with continuous transitions, 
$q=1$,2,3,4.  Except for the Ising case we have used the
topological rule.  The extension rule is used for the Ising
case. Figure \ref{fig:qpottw} is a log-log plot of
var$(f)^{1/2}$ vs.\/ the system size for the two-dimensional
Potts models with continuous transitions.  The figure shows that
the $f$ distribution becomes narrow as a power of the system
size $L$.  Fitting the last five data points for each $q$ to the
form, var$(f)^{1/2} \sim L^{-b(q)}$ yields $b(1)=.71(1)$,
$b(2)=.46(2)$, $b(3)=.30(2)$ and $b(4)=.23(1)$.  For percolation
this result is close to the expected scaling, $b(1)=1/\nu(1)$. 
For invasion percolation it is believed that the full $f$
distribution scales with $L^{-1/\nu(1)}$.       For the other
values of $q$, $b(q)$ is much smaller than $1/\nu(q)$ and
decreasing with $q$.  We do not yet understand the finite-size
scaling of the $f$ distribution.  

\subsection{First order transitions in {P}otts models}
\subsubsection{Three-dimensional 3-state Potts model}

We first discuss results for the topological and extension
stopping rules for the three-dimensional 3-state  Potts model. 
This model has a weak first order transition.   Fig.\
\ref{fig:pm3d-3} shows the mean and median values of $f$ vs.
$L^{-1}$ for system sizes between 10 and 70 for the topological
stopping rule and between  10 and 110 for the extension stopping
rule. Data was taken from samples of $10^4$ MCs for each lattice
size  up to $L=70$ and between 3000 and 6000 MCs for the larger
sizes.  In the case of the topological stopping rule, a fit of
the mean to a function of the form $c_{0} + c_{1} L^{-1/2} +
c_{2} L^{-1} + c_{3} L^{-2}$  gives $c_{0}$ = 0.4232. This is in
good  agreement with results obtained with other methods
\cite{Sch}. With the topological stopping rule, the median shows
no finite-size effects within the error, as is the case for the
three-dimensional Ising model. It can be fitted to a practically
horizontal line, which extrapolates to 0.4228. Again the median
seems to be the better choice if predictions about the infinite
system are to be made, although it tends to be more noisy. As in
the case of the two-dimensional Ising model, the difference
between the mean and the median results from a tail in the
distribution of $f$ towards $f=1$.

For the extension stopping rule, the median enters a flat region
starting at about $L=40$. The arithmetic mean of the last seven
data points is .42336. Current values of $T_{c}$ \cite{Sch}
agree in the first four  digits, and so does the value that we
obtained  from the median in this way.  Again, the mean is above
the median and starts to approach it only for very  large system
sizes ($L \geq 80$).

\subsubsection{Mass rule for first-order transitions} The above
results show that spanning rules may be used to accurately
locate a weak first-order transition however, they perform
poorly for strong first-order transitions (large values of $q$).
The difficulty is that the $f$ distribution becomes increasingly
broad and asymmetric with a tail extending toward $f=1$.  We
believe that the tail in the $f$ distribution is related to the
way the spanning condition is met for strong first-order
transitions.  The spanning cluster is a nearly linear object
which extends across the system in a background of small
clusters whose size is presumably the correlation length.  For
large values of $q$ the spanning cluster is very narrow
(somewhat like a river running through a terrain of small
clusters).  This observation is consistent with the increase in
$p(T_c)$ with increasing $q$.  In addition, for large $q$, there
are severe bottlenecks; one of only a few bonds must be occupied
to meet the spanning condition. This leads to a broad $f$
distribution with a tail toward $f=1$.  Although we believe the
$f$ distribution becomes sharp for large $L$ the convergence is
very slow.  An additional difficulty in using either of the
spanning rules for strong first-order transitions arises from
the fact that the spanning cluster is nearly reproduced in
successive Monte Carlo steps so that the autocorrelation time is
large.
 
These problems can be avoided with the mass stopping rule.  At
the phase coexistence temperature $T_{c}$, the  magnetization
may take any value from $0$ to $m_{l}$ where $m_{l}$ is the
magnetization of the pure low temperature phase. Thus we expect
$f$ to approach $p(T_c)$ for every $m$ between 0 and $m_{l}$. At
$m_{l}$ we expect the derivative of $f(m)$ to jump to a finite
value as the systems enters the ordered phase with $T<T_{c}$. 

Figure \ref{fig:pm2d-10}  shows a plot $\langle f \rangle$ vs.\
$m$ for the two-dimensional 10-state Potts model, obtained from
the mass  rule for system sizes 50, 100 and 200. Data was
obtained after an equilibration of 1000 MCs from a sample of
2$\times 10^4$ to 5$\times 10^4$ MCs.  The dashed line denotes
the exact value of $p(T_c)=\sqrt{q}/(1+\sqrt{q})$.  It is clear
that the crossing point of these curves (near $m=0.6$) for
different system sizes provides an accurate estimate of
$p(T_c)$. Furthermore, the curves become increasing flat for
large $L$ and presumably converge (non-uniformly) to $p(T_c)$. 
The value of $m_l$ can be estimated from the largest value of
$m$ for which $\langle f \rangle = p(T_c)$.  From data for
$L=50$, 100,150, 200 and 500 (the data for 150 and 500 is
omitted from the plot for clarity) we find convergence to the
value, $m_l = 0.8544$.

For small values of $m$ we find a region which moves
increasingly toward zero where $f$ is significantly greater than
$p(T_c)$.  The non-monotonicity of $f$ as a function of $m$
occurs only for those Potts models with first-order transition
(compare Fig.\ \ref{fig:pm2d-2-4-5}).

In the inset of Fig.\ \ref{fig:pm2d-10} the standard deviation
of $f$ is plotted vs.\ $1/L$. This quantity appears to vanish as
$L \rightarrow \infty$ though at a rate that depends on $m$. The
data is consistent with our belief that the $f$ distribution
approaches a delta function at $p(T_c)$ for any $m < m_{l}$. Any
mixture of low and high temperature phases in the coexistence
region can be sampled by fixing the ratio, $m/m_{l}$. This
argument can be confirmed by looking at the energy per spin. 
Let $n_{l}$ be the fraction of the system that is in the low
temperature  phase and assume that it is proportional to $m$. If
$\varepsilon_{l}$  ($\varepsilon_{h})$ is the energy per spin of
the pure low (high)  temperature phase, we expect the energy per
spin $\varepsilon$ to behave like  \begin{equation}  \langle
\varepsilon \rangle =   (m/m_{l}) \varepsilon_{l} + (1-m/m_{l}
)\varepsilon_{h} + \varepsilon_{sf}A/N  \label{eq:enansatz} 
\end{equation}  at the transition point. The third term includes
the interfacial energy  per unit area $\varepsilon_{sf}$ and the
area $A$ and should vanish with  system size like $1/L$.

Fig.\ \ref{fig:energypm2d-10} shows $\langle \varepsilon \rangle$
vs.\ $m$ for $L$ = 50, 100 and 200 from the same runs as the
data of Fig.\ \ref{fig:pm2d-10}.  Error bars obtained with the
jack-knife method are smaller than the symbols. There is a large
region from small $m$ to about $m_{l} = 0.855$ where the energy
is described by a line with negative slope plus a correction that
vanishes as $L$ becomes large. Again, this statement is also
based on additional data for $L$ = 150 and 500.  This behavior is
in good agreement with  Eq.\ \ref{eq:enansatz}. At $m_{l}$ the
energies for  different $L$ collapse, as there is no interfacial
energy left, and the systems enter the ordered phase below
$T_{c}$, causing the energy to drop. Only for small $m$ the
energies leave the presumed curve due to finite  size effects. In
order to estimate the energy of the high temperature phase, we
fitted the $L=200$ data points in the range  $0.05 \leq m \leq
0.5$ to a function of the form  $c_{0} + c_{1}m + c_{2} m^{2}$,
that gives $\varepsilon_{h}=c_{0}= -0.969$, which  agrees well
with the exact result $-0.9682 \ldots$ \cite{Bax}.  Evaluating
the data for the biggest system used, $L = 500$ at $m_{l}$ gives
$e_{l}$ = -1.661, which is close to the exact value $-1.6642
\ldots $ as well.

There are few computational methods for reliably distinguishing
the order of a phase transition \cite{LeKo}.  The 3-state Potts
model in three dimensions for example at the phase transition
point in many respects behaves  just like a second-order
transition. If the magnetization rule is used, however, even
weak first-order transitions seem to behave differently than
continuous ones. We observed that $f$ is a non-monotonic
function of $m$ for first-order transitions.  Figure
\ref{fig:pm2d-2-4-5} shows the median value of $f$ vs.\ $m$ for
Potts models for several values of $q$. The known values of
$p(T_c)$ are marked by dashed  lines. Note that the curves are
monotone increasing for models with second-order transitions,
($q =$ 2, 4) and non-monotonic for models with first-order
transitions  ($q=$ 5, 6 and 10). Non-monotonicity is also found
for the three-dimensional 3-state Potts model, see Fig.\
\ref{fig:pm3d-3fm}.  It should be noted that both the
three-dimensional 3-state and two-dimensional 5-state Potts
models have extremely weak first-order transitions so that this
criterion is quite sensitive. It is useful even if the
correlation length is larger than the system size. 

\subsection{Dynamics of the IC algorithm} \label{secauto}

In this section we study the dynamic properties of the IC
algorithm. The normalized autocorrelation function of an
observable $A$ is defined by, \begin{equation}
 \Gamma_A(t)=\frac{<A_0 A_{t}>-<A>^2}{<A^2>-<A>^2},
\end{equation} where $t$ is time in Monte Carlo steps. The three
sets of points in Fig.\ \ref{figd1} are the normalized
autocorrelation functions of the absolute value of the
magnetization $m$, energy $\varepsilon$, and fraction of
occupied bonds $f$ for the two-dimensional Ising model.
Numerical data were collected for the topological stopping rule
from a run of 10$^4$ MCs which was divided into 10 groups with
errors estimated by the jack-knife method. In a few steps all
three autocorrelation functions have nearly vanished. The
autocorrelation functions of $f$  and $\varepsilon$ display a
negative overshoot on the first step which becomes larger for
larger system sizes.  

The integrated autocorrelation time $\tau_A$, which is required
for estimating the errors in measuring the observable $A$, is
defined by, \begin{mathletters} \begin{equation}
 \tau_A(w)=\frac{1}{2}+\sum_{t=1}^w\Gamma_A(t), \end{equation}
\begin{equation}
 \tau_A=\lim_{w\rightarrow\infty}\tau_A(w). \end{equation}
\end{mathletters} The integrated autocorrelation time determines
the size of the standard error in measuring $A$ according to,
\begin{equation} \label{eq:errtau}  \delta A= ( 2 {\rm var}(A)
\tau_A/N_{MC})^{1/2}  \end{equation} with var$(A)$ the variance
in $A$ and $N_{MC}$ the number of Monte Carlo steps. Figure
\ref{figd2} is a plot of $\tau_A(w)$ with $A=m$, $\varepsilon$
and $f$ as a function of $w$ for the two-dimensional Ising
model.  Similar  behavior was found for the three-dimensional
Ising model. The error for $\tau_A(w)$ was estimated by taking
the square root of the sum of the variances of $\Gamma_A(t)$'s
for $t\leq w$.  For all three observables, $\tau_A(w)$ reaches a
plateau in a few steps.  

Table \ref{table1} is a summary of the integrated autocorrelation
time for the two and three-dimensional Ising model at  $w=6$,
where all the $\tau$'s have saturated but still have relatively
small errors.  The values for $\tau_\varepsilon$ are compared
with results~\cite{Wolff} for the SW and Wolff single cluster
algorithm. The energy autocorrelation time is markedly smaller
for the IC algorithm than the other two cluster algorithm.
Furthermore,  $\tau_\varepsilon$ decreases for larger systems
while for the other cluster algorithms it increases.   $\tau_m$
appears to be independent of system size suggesting the
possibility that the dynamic exponent for IC dynamics is zero
for both two and three-dimensional Ising models.  

For the largest systems, $\Gamma_f(1)$ is close to $-0.5$ and,
as a result, $\tau_f$ is very small. Although the data is not
good enough to draw clear conclusions, it appears that $\tau_f$
approaches zero as a power, $L^{z_f}$ with $z_f \approx -1$ .
The anticorrelation in $f$ and $\varepsilon$ means that these
quantities can be accurately estimated in a small number of
Monte Carlo steps.  Indeed, for averaging these variables, the
IC algorithm is better than performing independent sampling from
the invariant IC measure. For example, Eq.\ \ref{eq:errtau} and
the behavior of var$(f)$, implies that $\delta f\sim
L^{-a}/\sqrt{N_{MC}}$ where $a \approx 1.3$.

Table \ref{table2} shows results for the integrated
autocorrelation time for the 3 and 4-state Potts models.  We
find again that the IC algorithm is much faster than the SW
algorithm though for $q=3$ and 4 there appears to be some
critical slowing.  Based on Table \ref{table2} we can obtain
estimates for the dynamic exponet for the magnetization,
$z_m(3)=.28$ and $z_m(4)=.63$.  Note however that these values
are less than the Li and Sokal~\cite{LiSo89} bound for the
dynamic exponent for the SW algorithm ($z \geq \alpha/\nu$).  

\section{Summary}

The invaded cluster method comprises a class of algorithms for
sampling equilibrium spin systems.  Because cluster growth is
controlled by a spanning rule rather than the temperature, the
method is able to simulate the phase transition point without
{\it a priori} knowledge of the phase transition temperature.
The transition temperature is, instead, an output of the
algorithm.  We have demonstrated this numerically for
Ising/Potts models in two and three dimensions.  

We may also use parameterized stopping rules to explore either
the coexistence region of discontinous transitions or the
critical region near a continuous transition.  For these rules
we specify a quantity such as the energy or susceptibility and
learn the corresponding temperature.  Using the mass rule we
have been able to sweep through the coexistence region of
first-order transitions and to obtain quantities such as the
energy of the high and low temperature coexisting phases.  The
behavior of the effective transition temperature with the mass
parameter apparently yields a very sensitive method to
distinguish continuous from discontinuous transitions.

The invaded cluster algorithm is very similar to the
Swendsen-Wang algorithm except that the occupied bonds are
determined by a stopping rule rather than the temperature.  We
argued that this leads to a feedback mechanism that forces the
system to the desired equilibrium state much faster than is the
case for the Swendsen-Wang algorithm.  The consequence is that
the algorithm is extremely fast. Measured  autocorrelation
times  are less than unity and  decrease with system size for
the energy and estimated critical temperature. The magnetization
integrated autocorrelation time is constant for the two and
three-dimensional Ising models but grows slowly for the 3 and
4-state two-dimensional Potts models.  We speculate that the
invaded cluster algorithm applied to Ising critical points has
no critical slowing down.  For this reason and because there is
no need to know the transition temperature, we believe the IC
method will prove to be the most efficient approach for high
precision measurements of critical properties.       

Although we have tested the algorithm in a number of setting and
supplied non-rigorous arguments for its validity much work
remains to be done in understanding the method and putting it on
a firm footing.  It is important to prove that the IC ensemble
is equivalent to the usual statistical mechanics ensembles. We
would also like to understand the finite-size scaling properties
of the IC ensemble since these differs from our naive expectation
in some cases.  

In this paper we have confined our attention to Potts models
however the method is much more broadly applicable.  In a future
paper~\cite{ChMaun} we will show how to use the approach for a
variety of discrete spin systems such as the Ashkin-Teller
model.  Similary, the embedding approach described by Wolff can
be used to generalize the method to $O(n)$ models.  

This work was supported by NSF Grants DMR-93-11580, DMR-95-0013P
and DMS-93-02023.

\section*{Note added} After completing this research we received
an interesting preprint~\cite{LiGl} that describes a ``fixed
cluster'' algorithm.  This algorithm uses a stopping rule based
on the extent $l$ of the largest cluster.  However, in contrast
to the fixed parameter rules used here, $l$ does not correspond
to a thermodynamic quantity. 

 \newpage

\begin{table}  
\caption{Integrated autocorrelation times for
two- and three-dimensional Ising models for the SW, Wolff and IC
algorithms. Results for the IC algorithm are measured at time
step $w=6$.}    
\label{table1}  
\begin{tabular}{crlcccc} d & L
&$\tau_{\varepsilon,SW}$\tablenotemark[1] &
$\tau_{\varepsilon,Wolff}$\tablenotemark[1]
          & $\tau_{\varepsilon,IC}$ & $\tau_{m,IC}$ &
$\tau_{f,IC}$ \\ \tableline 2 & 32 & 4.13(4) & 1.80(1) & .51(3)
& .88(3) & .19(3) \\ 2 & 64 & 4.92(8) & 2.23(3) & .42(2) &
.78(3) & .11(2) \\ 2 & 128& 6.00(8) & 2.69(4) & .42(3) & .80(3)
& .07(3) \\ 2 & 256&	 & 3.17(8) & .37(3) & .81(3) & .06(3) \\
  &    &	 &	 &	  &	   &	    \\ 3 & 16 & 5.6(1)	 & 1.36(2) &
.35(2) & .65(2) & .09(2) \\ 3 & 24 & 6.8(1)  & 1.50(3) & .27(2)
& .62(3) & .07(2) \\ 3 & 32 & 7.8(3)  & 1.72(4) & .25(2) &
.65(2) & .05(2) \\ 3 & 48 & 9.9(4)  & 1.90(6) & .19(2) & .66(3)
& .02(3) \\ \end{tabular} \tablenotetext[1]{Ref.\
\cite{Wolff89b}.} 
\end{table}  
\begin{table}
\caption{Integrated autocorrelation times for two-dimen\-sional,
3 and 4-state Potts models for the SW and IC algorithms. Results
for $\tau_{\varepsilon,IC}$ and $\tau_{f,IC}$ for IC dynamics
are measured at time step $w=6$ while the time step $w_m$ for
$\tau_{m,IC}$ is shown in the last column. Results for  SW
dynamics are for sizes 128 and 256 rather than 120 and 250.} 
\label{table2}  \begin{tabular}{clccccl} q & L
&$\tau_{\varepsilon,SW}$\tablenotemark[2] &
$\tau_{\varepsilon,IC}$ & $\tau_{f,IC}$&$\tau_{m,IC}$&$w_m$ \\ 
\tableline 3 & 120 					&30.3(1.2)  & .73(3) & .08(3)&	1.40(4)&	
6 \\ 3 & 250					 &39.6(1.7)  & .59(2) & .06(2)&	1.73(5)&		11\\
3 & 500	     &           & .52(2) & .06(3)&	2.08(6)&		15\\
  &    	     &           &        &       &					   &			 \\ 4 &
120					 &115.7(6.1) & 1.23(2)& .11(3)&	2.97(7)&		16\\ 4 &
250					 &232.0(24.6)& 1.10(3)& .09(2)&	4.61(10)&	27\\ 4 &
500	     &           & 0.88(3)& .05(2)&	7.31(16)&	57\\
\end{tabular} \tablenotetext[2]{Ref.\ \cite{LiSo89}.} \end{table}

\begin{figure}
\caption{$\langle f \rangle$ vs.\ 
$L^{-1/\nu}$ for the three-dimensional Ising model using
different stopping rules. The infinite volume estimate of
$p(T_c)$ from Ref.\ \protect\cite{FeLa} is shown on the vertical
axis.}  \label{fig:spancompare} \end{figure}

\begin{figure}
\caption{Double logarithmic plot of
$\langle f \rangle_{t} - \langle f \rangle_{e}$ vs. $L$ for the
three-dimensional Ising model.}  \label{fig:fdiff} \end{figure}

\begin{figure} 
\caption{Double logarithmic plot of the
distribution of cluster sizes $n(s)$ for the three-dimensional
Ising model. The inset shows a double logarithmic plot of the
average size of the largest cluster.} \label{fig:csdist3d-2}
\end{figure}

\begin{figure} 
\caption{The mean and the median of $f$  vs.\
$1/L$ for the two-dimensional Ising model. The solid line shows
a linear fit to the median and the exact infinite  volume value
is shown on the vertical axis. The inset shows the standard
deviation of the $f$ vs. $1/L$.  A least squares fit to the form
$c_{0} + c_{1}L^{-1/2} + c_{2}L^{-1}$ (solid  line) suggests
that the distribution becomes sharp in the infinite  volume
limit.}   \label{fig:f2dising} \end{figure}

\begin{figure} 
\caption{$\langle \varepsilon \rangle$
vs. $1/L$ for the two-dimensional Ising model.  The solid line
is a fit to the form
$\varepsilon_{0}+\varepsilon_{1}L^{-1}+\varepsilon_{2}L^{-2}$
and the  exact infinite volume result is shown on the vertical
axis.  The inset shows var$(\varepsilon)N$ vs.\ $L$ with a
linear fit through the data.}  \label{fig:energypm2d-2}
\end{figure}

\begin{figure}  
\caption{The mass rule
applied to the two-dimensional Ising model. The upper graph
shows $\langle f \rangle$ vs.\ $m$. In the lower
graph the same data is scaled as described in the text.}  
\label{fig:pm2d-2} \end{figure}

\begin{figure} 
\caption{Double logarithmic plots of 
$|f-p(T_c)|$ vs.\ system size $L$ for
 the two-dimensional $q$-state Potts models. Exact values of
$p(T_c)$ are used.} \label{fig:qpottf} 
\end{figure}

\begin{figure} 
\caption{Double logarithmic plots of the
standard deviation of $f$, var$(f)^{1/2}$ vs.\ system size $L$
for the two-dimensional $q$-state Potts models.}
\label{fig:qpottw}
\end{figure}

\begin{figure} 
\caption{The mean and median of $f$ for the
three-dimensional 3-state Potts model. The upper graph shows
results from the topological stopping rule, the solid line is a 
linear fit to the median. The lower graph shows results from the
extension stopping rule.}  \label{fig:pm3d-3} \end{figure}

\begin{figure} 
\caption{$\langle f \rangle$ vs.\ $m$ for the
two-dimensional 10-state Potts model using the mass rule.   The
dashed line marks the exact value, $p(T_{c})$. The inset shows
the standard deviation of $f$ vs.\ $1/L$ for $m = 0.4$ (squares)
and $m = 0.85$ (triangles) .}   \label{fig:pm2d-10} \end{figure}

\begin{figure}
\caption{$\langle \varepsilon \rangle$
vs.\ $m$ for the  two-dimensional 10-state Potts model using the
mass rule for several lattice sizes $L$.  The solid line is a
fit to the form $c_{0} + c_{1}m + c_{2} m^{2}$ for the $L=200$
data points with $0.05 \leq m \leq 0.5$, the intercept with the
$m = 0$ axis is our estimate for $\varepsilon_{h}$, the energy
of the high temperature phase at the transition .}  
\label{fig:energypm2d-10} \end{figure}

\begin{figure}
\caption{The median of $f$ vs.\  $m$
for the two-dimensional Potts models using the mass rule with
$q$ = 2, 4, 5, 6, and 10 and $L=200$. The exact value of
$p(T_{c})$ for each $q$ is shown by a dashed line.} 
\label{fig:pm2d-2-4-5} \end{figure}

\begin{figure}
\caption{$f$ vs.\ $m$ for the
three-dimensional 3-state Potts
model.} \label{fig:pm3d-3fm} \end{figure}

\begin{figure} 
\caption{Autocorrelation functions of the
magnetization $m$, energy $\varepsilon$ and occupation fraction
$f$ vs.\ time step $t$ for the two-dimensional Ising model for
size, $L=256$.} \label{figd1} \end{figure} 
 
\begin{figure}
\caption{Integrated autocorrelation times for $m$, $\varepsilon$
and $f$ vs.\ integration time, $w$ for the two-dimensional Ising
model for size, $L=256$.} \label{figd2} \end{figure}
\end{document}